\documentstyle[aps,prl,epsfig,amssymb]{revtex}
\begin{document}

\draft

\title{Disorder induced hexagonal-orthorhombic transition in
Y$^{3+}_{1-x}$Gd$^{3+}_{x}$MnO$_3$}

\author{Jan-Willem G. Bos, Bas B. van Aken and Thomas T. M. Palstra}
\address{Solid State Chemistry Laboratory, Materials Science Centre, University of
Groningen, Nijenborgh 4, 9747 AG Groningen, the Netherlands}

\date{\today}
\twocolumn[\hsize\textwidth\columnwidth\hsize\csname@twocolumnfalse\endcsname

\maketitle

\begin{abstract}
We show that the transition in AMnO$_3$ from the orthorhombic perovskite phase
to the hexagonal phase is promoted by inducing disorder on the A-site. The gap
between the orthorhombic and the hexagonal phase is widened for disordered,
mixed yttrium-gadolinium manganite samples. At the cost of the orthorhombic
phase a two phase region emerges. The phase separation exhibits very unusual
thermodynamical behaviour. We also show that high pressure synthesis favours the
orthorhombic phase. YMnO$_3$ is formed in the orthorhombic phase at 15 kbar.
\end{abstract}


]

\section{Introduction}
In the search for new composition-properties relations ABO$_3$ compounds have
attracted a lot of attention. The perovskite materials, ABO$_3$, have been
researched extensively because this structure forms the basis for interesting
physical properties such as high $T_c$ superconductivity\cite{Bed86} and
colossal magnetoresistance\cite{Jin94a}. Non-perovskite AMnO$_3$, with A = Y,
Ho,...,Lu, attracted renewed interest, due to their ferroelectric
properties\cite{Ber63a}. These hexagonal AMnO$_3$\cite{Yak63} have a basically
different structure than most ABO$_3$ compounds, that are distorted perovskites.
These properties arise due to the strong correlation of the $3d$ electrons with
the O $2p$ orbitals.

In this paper, we report the transition from the orthorhombic perovskite (o) to
the hexagonal (h) phase by changing the ionic radius of the A ion and by high
pressure synthesis. The effect of the ionic radius on the transition is studied
by partially replacing Y by Gd ions in YMnO$_3$. The resulting phase diagram
leads us to discuss the effect of disorder, in terms of the ionic radius
variance, on the stability of the hexagonal and orthorhombic phases.

The basic building block of the perovskite is an oxygen octahedron with a
transition metal, B, in its centre. The A ions, usually lanthanides or alkaline
earth metal ions, occupy the holes between the octahedra, that form a 3D corner
shared network. In this picture B is sixfold and A is 12-fold coordinated. Most
perovskites have a distorted structure, derived from this building block. The
distortions have various origins, including a ferroelectric transition for B a
$d^0$ transition metal ion like Ti$^{4+}$ \cite{von46}. The most common
distortion originates from the relative small radius of the A ions compared with
the holes between the octahedra. This results in a cooperative rotation of the
octahedra known as the GdFeO$_3$ distortion\cite{Gla72}. While the structure is
interesting in its own right, it has also large effects on the physical
properties. It is well documented that the physical properties depend strongly
on the magnitude of the structural distortions. An overview for the manganites
is given in Ref.'s~\cite{Ram97,Coe99}.

The magnitude of the GdFeO$_3$ distortion depends strongly on the tolerance
factor, $t$:
\begin{equation}
 t=\frac{r_{A^{3+}}+r_{O^{2-}}}{\sqrt2(r_{B^{3+}}+r_{O^{2-}})},
\end{equation}
where $r_X$ is the radius of the X ion. The tolerance factor gives the relation
between the radii of ions A, B and O in a ideal cubic perovskite. For $t=1$ the
size of the lanthanide is exactly right to compose the cubic perovskite system.
For Mn$^{3+}$, $r_{Mn^{3+}}=0.645$ \AA\ and $r_{O^{2-}}=1.42$ \AA\ this yields a
ionic radius $r_{A^{3+}}=1.50$ \AA, where the largest lanthanide, La, has a
radius of 1.22 \AA. The corresponding tolerance factor $t=0.90$ indicates a
large distortion for LaMnO$_3$. With increasing atomic number, the lanthanide
radius decreases and thereby the distortion increases. For the manganites, the
tolerance factor is conventionally regarded as the factor controlling the
boundary between the hexagonal and orthorhombic structures. The orthorhombic
perovskite phase is stable for $t>0.855$, corresponding to $r_A\geqslant
r_{Dy}$\cite{Yak55}. For $t<0.855$, $r_A\leqslant r_{Ho}$, the hexagonal phase
prevails \cite{Yak63}. Yttrium, although not in the lanthanide series, behaves
chemically identical and its radius falls between dysprosium and holmium. An
overview on the ionic radii and tolerance factors of some relevant compounds is
given in Table~\ref{overview}.

\begin{table}[htb]
 \centering
 \caption{Ionic radii and tolerance factors for relevant compounds.}
 \begin{tabular}{ccc}
 Compound & tolerance factor & ionic radius (\AA) \\
 LaMnO$_3$ & $0.902$ & $1.215$ \\
 GdMnO$_3$ & $0.866$ & $1.109$ \\
 DyMnO$_3$ & $0.857$ & $1.083$ \\
 YMnO$_3$ & $0.854$ & $1.075$ \\
 LuMnO$_3$ & $0.840$ & $1.032$ \\
 \end{tabular}
 \label{overview}
 \end{table}

\begin{figure}[htb]
   \centering
   \includegraphics[height=5.6cm,width=8cm]{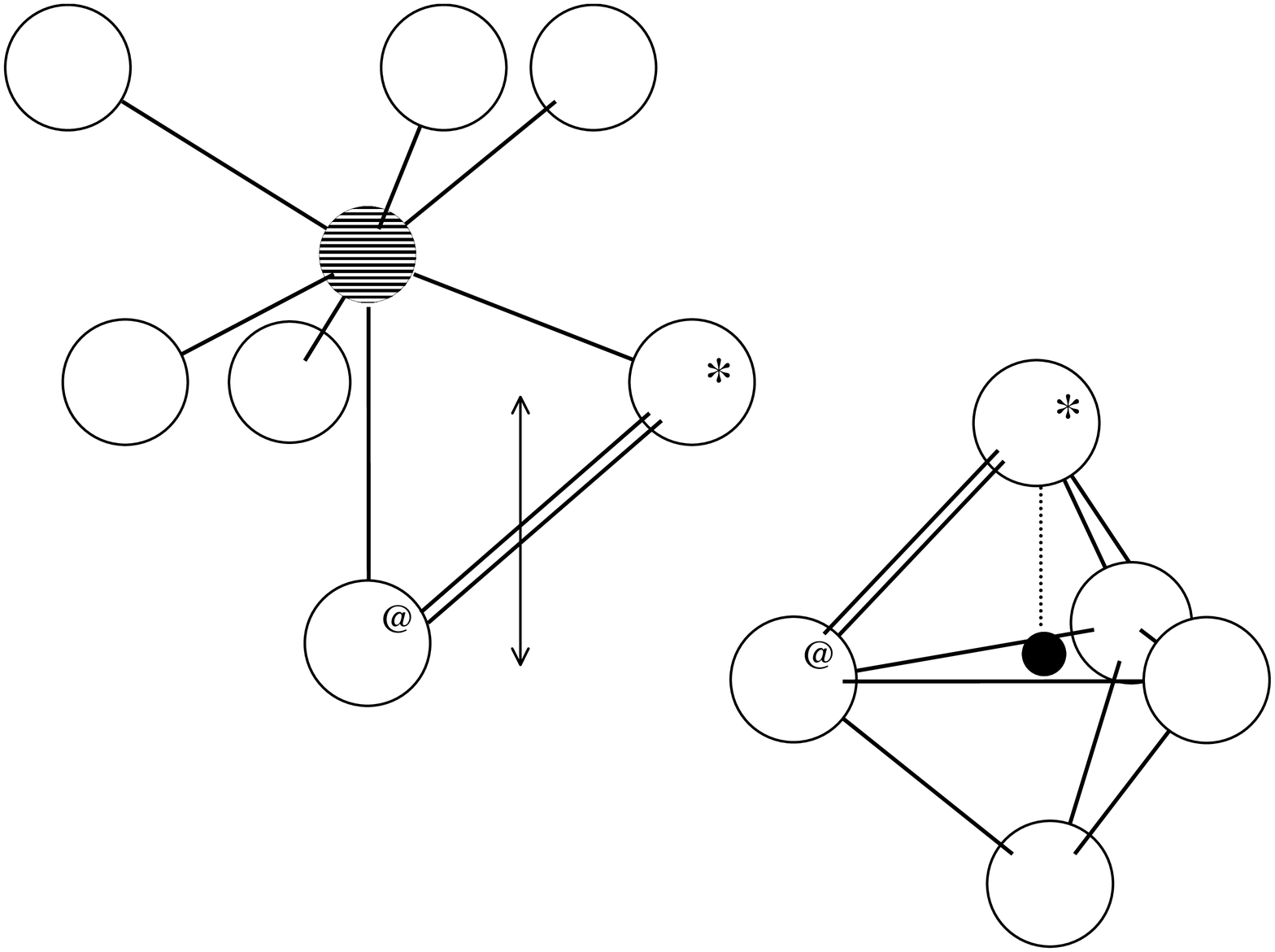}
   \caption{Sketch view of the local environment, showing AO$_7$, left side, and MnO$_5$,
right side. The arrow indicates the distance between two oxygen planes. The
dashed line indicates the Mn-O$_{ap}$ distance. Atoms marked with "*" and with
"@" are identical, the double line indicates the shared edge.}
   \label{poly}
\end{figure}

The high temperature phase of the hexagonal AMnO$_3$ consists of 8-fold
coordinated A ions in bicapped antiprisms. Trigonal bipyramidal holes are formed
between two layers of face-sharing antiprisms by the edges of the capping
oxygens with the antiprisms. The capping oxygens of two adjacent layers are
located on the same $ab$ plane. Half of the bipyramidal holes are occupied by
Mn. The apical oxygens of the MnO$_5$ bipyramid are also the oxygens that make
up the antiprism. The two polyhedra are sketched in Fig.~\ref{poly}, where the
shared edge is shown. The Mn-O$_{ap}$ distance is thus equal to the distance
between the antiprism oxygen layer and the capping oxygen layer. The steric
hindrance of the Mn restricts this layer separation and therefore increases the
A-O$_{ap}$ bond length. Thus, the eightfold co-ordination is not uniform. The
two apical oxygens have slightly larger bond lengths. Furthermore, the structure
is unstable against a ferroelectric distortion at lower temperatures. The apical
oxygens move in such a way that one bond becomes 'normally' short, while the
other becomes about 1 \AA\ larger. The asymmetric A environment is the main
reason for the ferroelectric behaviour. As we have two lanthanide positions in
$P6_3cm$, we have two non-equivalent, although similar, dipole moments. Four out
of the six moments per unit cell point upwards, the other two downwards.

Although the hexagonal phase of YMnO$_3$ at ambient conditions is the
thermodynamically stable phase, there are several ways to obtain orthorhombic
YMnO$_3$. Using thin film growth, an appropriate substrate will force the
coherent growth of the orthorhombic phase \cite{Sal98}. Synthesis routes via
organic precursors and low reaction temperatures yield the orthorhombic phase
\cite{Bri97}. And last, high pressure synthesis favours the orthorhombic phase,
because it has a higher density \cite{Wai67}.

\section{Experimental}
Polycrystalline ceramic samples of AMnO$_3$, where A is a mixture of Y and Gd,
have been synthesised using regular solid state synthesis at ambient pressure.
Starting materials were Y$_2$O$_3$, Gd$_2$O$_3$ and MnO$_2$. Stoichiometric
amounts corresponding to formulae which range from pure YMnO$_3$ to pure
GdMnO$_3$ were weighted and wet-mixed using acetone as liquid medium. The
pressed pellets were sintered for 24 hours at 1250$^\circ$ and for 24 hours at
1400$^\circ$.

High pressure experiments were carried out both on the mixture of oxides and on
as-prepared samples. X-ray powder diffraction patterns were identical for both
methods. The high pressure high temperature piston cylinder apparatus is a Depth
of the Earth Quickpress 3.0, with an experimental range up to 25 kbar and
2100$^\circ$C \cite{Quickpress}. The lower pressure limit is $\sim1$ a 2 kbar.

The sample environment is a complex set-up, including a graphite furnace and the
pressure medium. Care has to be taken to prevent contamination from the graphite
resistance furnace or any of the other materials, \emph{e.g.} Al$_2$O$_3$ or
NaCl, in the sample assembly. Therefore, a small amount of powder $\sim0.5$ g
was encapsulated in a Pt capsule. The capsule consisted of a tube (diameter 4 mm
and height 6 mm) and two pre-shaped lids, which were welded together using a
small welding apparatus. Recovery of the pellet from the sample assembly is
improved by the Pt container.

X-ray diffraction patterns were recorded using a Bruker-AXS D8 powder
diffractometer, with primary and secondary monochromator, using Cu $K_\alpha$
radiation. Patterns were analysed for phase determination using the evaluation
software EVA \cite{EVA}, the Powder Diffraction File \cite{PDF} and the
Inorganic Crystal Structure Database \cite{ICSD}. Patterns of non-contaminated
samples, containing only the hexagonal and orthorhombic phases, were used in the
Rietveld refinement using TOPAS R \cite{TOPASR}. Rietveld refinements included
lattice parameters, zero point correction and the ratio between the two phases.
Atomic positions were assumed to be constant using the positions determined by
single crystal x-ray diffraction on YMnO$_3$ \cite{Van01a}. The ratio Y:Gd was
fixed at the nominal composition.

\begin{table}[htb]
  \caption{Ionic radii, tolerance factors and variance of the studied
Y$_{1-x}$Gd$_{x}$MnO$_3$ samples.}
  \begin{tabular}{lccr}
    $x$ & $r_A$ (\AA) & $t$ & $\sigma^2$ ($10^{-6}$\AA) \\
    0    & 1.075 & 0.854 & 0 \\
    0.06 & 1.077 & 0.855 & 68 \\
    0.19 & 1.081 & 0.857 & 176 \\
    0.25 & 1.084 & 0.857 & 217 \\
    0.31 & 1.086 & 0.858 & 248 \\
    0.38 & 1.088 & 0.859 & 271 \\
    0.5  & 1.092 & 0.860 & 289 \\
    1    & 1.109 & 0.866 & 0
  \end{tabular}
  \label{samples}
\end{table}

\section{Results and discussion}
\begin{figure}[htb]
   \centering
\includegraphics[width=80mm,height=59.7mm]{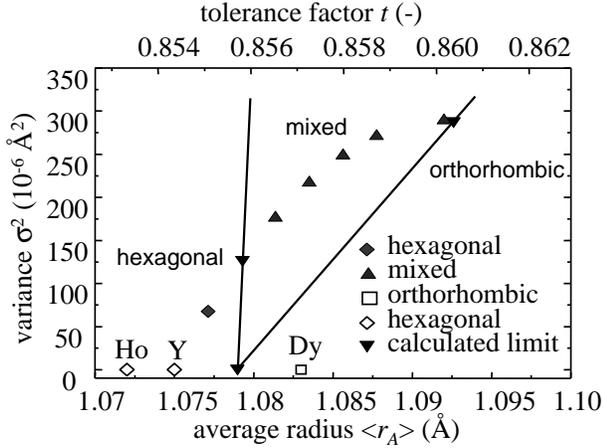}
   \caption{Phase diagram of Y$_{1-x}$Gd$_x$MnO$_3$ as a function of $r_A$, or $t$, and the
variance, $\sigma^2$. Diamonds indicate hexagonal phase, triangles mixed phase
and squares orthorhombic phase. The drawn lines are estimates of the phase
boundaries as explained in the text. The end member GdMnO$_3$ has $r_A=1.109$
\AA.}
   \label{varrad}
\end{figure}

In Fig.~\ref{varrad}, we present a phase diagram of Y$_{1-x}$Gd$_{x}$MnO$_3$ as
a function of $r_A$ and its variance $\sigma^2$ of samples given in
Table~\ref{samples}. The variance $\sigma^2$ is given by,
\begin{equation}
  \sigma^2= \sum_1^n x_i(r_i-\langle r_A\rangle)^2
\end{equation}
The phase diagram can be divided in three regions. Low $r_A$ compounds, $\langle
r_A \rangle<1.078$, are hexagonal. Large $r_A$ and small $\sigma^2$ compounds
are orthorhombic. The intermediate region shows both phases. The data for
HoMnO$_3$ and DyMnO$_3$ have been taken from the literature \cite{Yak63}.

By changing the value of $x$ we substitute Y by Gd, whereby $r_A$ increases
linearly. The tolerance factor of orthorhombic DyMnO$_3$ is equal to that of
Y$_{1-x}$Gd$_x$MnO$_3$, with $x=0.23$. Thus for $x\gtrsim0.23$ we expect to see
the orthorhombic phase. For smaller $x$ and $t$ the hexagonal phase is expected.
For $x=0$ and $x=0.06$ we indeed found the hexagonal structure. However, for
$0.19\leqslant x\leqslant0.38$ we do not observe a sharp transition to the
orthorhombic structure, but a mixture of the hexagonal and orthorhombic phases.
Only for $x=0.5$ an (almost) pure orthorhombic compound is found. The anomalous
behaviour of the mixed Y,Gd samples is best illustrated by focussing on the
sample with $x=0.25$, which has an almost identical tolerance factor as
DyMnO$_3$. Where the latter sample is orthorhombic, the former segregates in
both phases. The only difference between the two compounds is that one is
undoped and the other has a mixed lanthanide composition.

\begin{figure}[htb]
   \centering
\includegraphics[height=48mm,width=72mm]{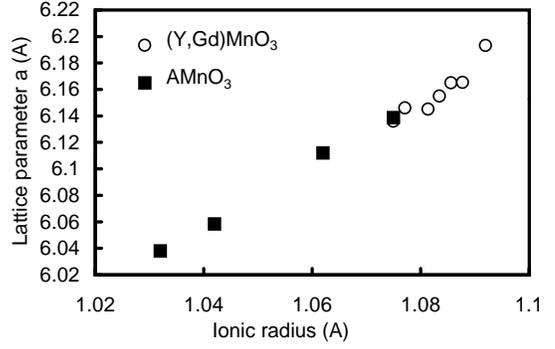}
   \caption{Lattice parameter $a$ of the hexagonal phase as a function of the Gd content
$x$.}
   %bounding box 60 305 380 517 w=320, h=212
   \label{Latpar}
\end{figure}

\begin{figure}[htb]
   \centering
\includegraphics[height=62.7mm,width=80mm]{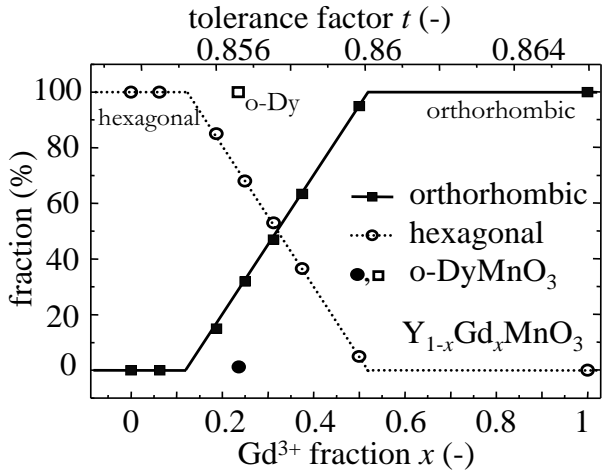}
\caption{Relative amounts of hexagonal, triangles, and orthorhombic, squares,
fractions as a function of the Gd content $x$. An open square is plotted for
DyMnO$_3$ at the corresponding value of the tolerance factor.}
   \label{Hexper}
\end{figure}

The relative amounts of orthorhombic and hexagonal fractions are determined by
Rietveld refinement of the powder diffraction data, see Fig.~\ref{Hexper}. We
can rule out element segregation of Y and Gd, since the continuous increase in
the lattice parameter $a$ of the hexagonal phase with increasing Gd
concentration indicates perfect mixing of the A ions. A linear increase in $a$
with $r_A$ is also observed for the single A cation h-AMnO$_3$ series
\cite{Yak63,Van01a,Van01f,Van01e}. Fig.~\ref{Hexper} shows that the orthorhombic
phase fraction increases linearly with the Gd fraction. This allows us to apply
the lever rule on the h-o transition
\begin{equation}
\frac{x-x_{hexa}}{x-x_{ortho}}=\frac{y_{ortho}}{y_{hexa}}
\end{equation}
where $x_{hexa}$ and $x_{ortho}$ are the boundary values for the respective
phases and $y_{ortho}/y_{hexa}$ the ratio of the two fractions. From the
observed ratios $y_{ortho}/y_{hexa}$ as a function of $x$, the boundary values
are derived. The boundary values derived from all five mixed phase samples are
plotted in Fig.~\ref{varrad} as inverted triangles.

We construct a preliminary phase diagram by assuming that the h-o transition at
$\sigma^2=0$ occurs halfway between hexagonal YMnO$_3$ and orthorhombic
DyMnO$_3$. The phase boundaries are drawn in Fig.~\ref{varrad} as straight lines
through the two calculated boundary values and the assumed $\sigma^2=0$
midpoint. This phase diagram can be described as follows. The phase line
associated with the upper $t$ limit of the hexagonal phase does not depend on
the variance. Slightly increasing $\sigma^2$ and $t$ results in the appearance
of a two phase region, consisting of both the hexagonal and the orthorhombic
phase. With increasing tolerance factor, the fraction of the orthorhombic phase
increases until the lower boundary limit for the orthorhombic phase is crossed.
This limit strongly depends on $\sigma^2$. The lower limit for the orthorhombic
phase increases from $r_A=1.078$~\AA\ at $\sigma^2=0$ to $r_A=1.093$~\AA\ at
$\sigma^2=\sigma^2_{max}$. Note that $r_A=1.093$~\AA\ corresponds with the ionic
radius of terbium, the second smallest lanthanide to form the perovskite
structure.

We have shown the dependence of the h-o transition on the average radius $r_A$
and $\sigma^2$. In the next section, the effect of high pressure experiments on
the h-o transition will be discussed.

\begin{figure}[htb]
   \centering
\includegraphics[width=80.0mm, height=68.6mm]{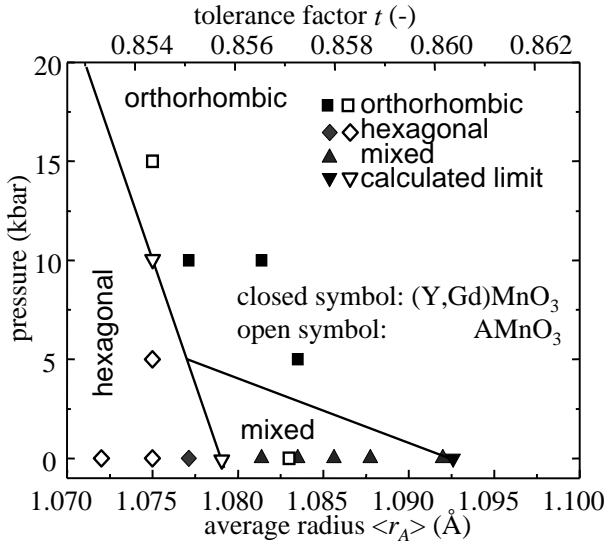}
   \caption{Pressure versus average radius phase diagram. Undoped samples are shown with
open symbols, doped Y-Gd samples with closed symbols. We never observed mixed
samples after high pressure synthesis. Schematic phase boundaries are drawn.}
   \label{tolrad}
\end{figure}

We applied high pressure and high temperature to convert or synthesise some of
the conventionally hexagonal samples in the orthorhombic state. Pressure
generally stabilises the most dense phase, in this case the orthorhombic
structure. First, we consider samples with $\sigma^2=0$. YMnO$_3$ is still
hexagonal at 5 kbar, but the orthorhombic phase is found using a pressure of 15
kbar. The data is shown in Fig.~\ref{tolrad}. The necessary pressure for the h-o
transition, less than 15 kbar, is much less than reported previously in the
literature \cite{Wai67}. The h-o transition phase line has been sketched in
Fig.~\ref{tolrad} by using midpoints. Extrapolating the pressure dependence of
the h-o transition yields a critical pressure of $\lesssim27$ kbar for
HoMnO$_3$. The error bar on this value, $\sim10$ kbar, is large because of the
sparse data points.

We also carried out high pressure experiments on the Y$_{1-x}$Gd$_x$MnO$_3$
compounds. The necessary pressure to induce the orthorhombic phase from the
phase mixture decreased with increasing $r_A$.

We note the following observations of the two-phase region:
\begin{enumerate}
\item The phase mixture is only observed for ambient presurre synthesis for samples
with $\sigma^2\neq0$.
\item We do not observe the mixed phase for any of the experiments at high pressure
($\geqslant 5$ kbar). We cannot exclude the presence of the mixed phase at low
pressures, but our experimental set-up is not well-suited for those pressures.
\item The lattice parameters of both phases in the two phase region indicate no
segregation into Y-rich and Gd-rich phases, see Fig.~\ref{Latpar}, which is
unconventional.
\item Literature reports that synthesis via organic precursors can result in a mixture
of hexagonal and orthorhombic phases for AMnO$_3$, $\sigma^2=0$, compounds
~\cite{Bri97}.
\end{enumerate}

These observations lead to the following conclusions. For $\sigma^2=0$ compounds
either the hexagonal or the orthorhombic structure is stable. Whereas low
temperature synthesis may yield mixed phase samples, a high temperature anneal
will convert the unstable phase. The hexagonal or orthorhombic phase will be
stable depending on the tolerance factor. However for $\sigma^2\neq0$ mixed
phase samples can be obtained for a broad range of tolerance factors. Even a
high temperature anneal retains the phase segregated state. Surprisingly, the
phase segregation is not accompanied by two limiting compositions, \emph{e.g}
Y$_{1-x}$Gd$_x$MnO$_3$ with $x=0.1$ and $x=0.5$. The continuous increase of the
lattice parameters of both the hexagonal and the orthorhombic phase throughout
the two phase region indicates that the composition of the hexagonal and the
orthorhombic state are the same in the mixed state. We have no explanation for
this unconventional form of phase segregation.

\section{Conclusions}

We have constructed phase diagrams for the h-o phases of AMnO$_3$, including the
effects of average ionic radius, hydrostatic pressure and variance. For
compounds with $\sigma^2\neq0$ a phase separation in the orthorhombic and the
hexagonal phase is found. The mixed region exists only at low pressures. We have
shown that at ambient pressure, this region expands towards higher values of the
average radius with increasing variance. The upper limit for the hexagonal phase
is not affected by an increase in the variance. We speculate that disorder,
introduced by a large variance or soft chemical synthesis routes, allows the
occurrence of the phase separation in the absence of other driving forces.
Suppressing the disorder by applying external pressure or annealing at high
temperatures prevents the existence of the phase separation. Pressure favours
the denser, orthorhombic phase, whereas thermal annealing promotes the hexagonal
phase.

%\ack
This work is supported by the Netherlands Foundation for the Fundamental
Research on Matter. (FOM).

%\bibliographystyle{c:/texmf/data/revtex/prsty}
%\bibliography{c:/texmf/data/revtex/bas}

\end{document}